\documentclass[10pt,letterpaper]{article}
\usepackage{opex3}
\usepackage{amsmath}

\begin{document}
\title{Adaptive Optics for \\ Fluorescence Correlation Spectroscopy}
\author{Charles-Edouard Leroux, Ir\`{e}ne Wang, Jacques Derouard\\ and Antoine Delon$^*$}
\address{Univ. Grenoble 1 / CNRS, LIPhy UMR 5588, Grenoble, F-38041, France}
\email{$^*$antoine.delon@ujf-grenoble.fr} 



\begin{abstract}

Fluorescence Correlation Spectroscopy (FCS) yields measurement parameters (number of molecules, diffusion time) that characterize the concentration and kinetics of fluorescent molecules within a supposedly known observation volume. Absolute derivation of concentrations and diffusion constants therefore requires preliminary calibrations of the confocal Point Spread Function with phantom solutions under perfectly controlled environmental conditions. In this paper, we quantify the influence of optical aberrations on single photon FCS and demonstrate a simple Adaptive Optics system for aberration correction. Optical
aberrations are gradually introduced by focussing the excitation laser beam at increasing depths  in
fluorescent solutions with various refractive indices, which leads to drastic depth-dependent bias in the estimated FCS parameters. Aberration correction with a Deformable Mirror stabilizes these parameters within a range of several tens of $\mu m$ into the solution. We also demonstrate, both theoretically and experimentally, that the molecular brightness scales as the Strehl ratio squared.
\end{abstract}



\section{Introduction}

Fluorescence Correlation Spectroscopy (FCS) belongs to a family of
methods based on the analysis of fluorescence fluctuations that aims at quantifying molecular dynamics and concentration (in the pM to $\mu$M range) \cite{FFM}. In its simplest form, FCS uses the temporal Autocorrelation Function (ACF) of the fluorescent signal collected at a single point
by the detector of a confocal microscope. The least-square fit of the ACF provides parameters such as the mean number of molecules in the FCS observation volume, $N$ and the diffusion time, $\tau_{D}$, through the observation volume \cite{RevueFCS, FCS-modèles}. The standard fitting approach uses an analytical form to model the ACF, which is derived assuming that the confocal Point Spread Function (PSF) is a 3D Gaussian function. For this reason, attention was initially paid to the discrepancy, in the aberration-free case, between the
actual confocal PSF and the ideal 3D Gaussian function and its consequences for FCS measurements \cite{hess02,Muller04,Huang01}.

FCS yields measurement parameters, $N$ and $\tau_{D}$, that scale with the observation volume, which makes it difficult to compare
measurements obtained with different samples (solutions, cells, tissues, \textit{etc.}) and in
different environments (temperature, substrates). Preliminary calibrations of the observation volume are usually performed with fluorophore solutions, with well known diffusion constant and  concentration \cite{FCS-calibration}. However, since optical aberrations depend upon optical alignments, refractive index of the solvent, observation depth, \textit{etc.}, these calibrations are unusable if not performed under the exact same conditions as the experiment of interest. The key role of refractive index mismatches in FCS samples has been stressed by Enderlein \textit{et al.} \cite{Derlinger,Muller09}, who introduced a new technique and the associated data processing to measure the observation volume while performing FCS measurements, using two overlapping FCS volumes. This so-called dual focus FCS is to some extend robust to optical aberrations, and is to our knowledge the best alternative to an adaptive aberration correction as the one introduced in this paper.

The extreme sensitivity of FCS to optical aberrations is commonly observed on scanning FCS systems, which provide simultaneous maps of dynamics parameters ($\tau_{D}$, $N$) and conventional images. The field dependence of optical aberrations can lead to a twofold increase in diffusion times and even more in molecular brightness, while the effects on the confocal image are barely noticeable \cite{Dross09,Ferrand09}. We therefore anticipate that optical aberrations can induce small deformations of the confocal PSF, to which FCS measurements are more sensitive than conventional confocal imaging. In particular, one can expect that the estimated number of molecules and diffusion time increase with optical
aberrations, while the molecular brightness should decrease. To study these effects we performed FCS acquisitions in various solutions of Alexa Fluor 647 at nM concentrations: i) pure water; ii) 50$\%$ (v/v)
aqueous glycerol; iii) and 70.4$\%$ (v/v) aqueous glycerol. The pure water solution was used to monitor the effect of well defined aberrations, introduced by the Deformable Mirror (DM), onto the FCS output parameters. The two later solutions have refractive indices 1.407 and 1.435, whereas the objective lens is optimized for water. These refraction index mismatches induce weak optical aberrations (up to 0.1 $\mu$m RMS), dependent upon the observation depth, that dramatically affect the FCS outputs ($e.g.$ the number of molecules is multiplied by more than a factor 4). The DM will be used in this case to correct the aberrations.

\section{FCS and aberration modeling}

FCS exploits the temporal ACF, $G(\tau)$, of the confocal fluorescent signal, $I(t)$, assumed to be stationary:

\begin{equation}
G(\tau) = \dfrac{\langle I(t)I(t+\tau)\rangle}{\langle I(t)\rangle^2}
\end{equation}

where $\langle . \rangle$ practically denotes a temporal averaging. The general procedure consists in fitting $G(\tau)$ with a given model that depends on a series of parameters. Since, in the present study, we only deal with solutions of a single fluorescence species, the standard FCS model for brownian diffusion is well-suited. The corresponding parameters are the number of molecules in the FCS observation volume, $N$, the diffusion time through this volume, $\tau_{D}$ and $S$, the structure parameter. Assuming that the confocal PSF is a 3D Gaussian function, these parameters can in turn be related to physical quantities, which are the concentration, the confocal PSF widths and the diffusion constant. As a matter of fact, the mean number of fluorescent molecules is related to the number density of fluorescent molecules, $C$, by:

\begin{equation}
N = C\times  \pi^{3/2}{w_{r}^{2}}{w_{z}}
\label{NCWrWz}
\end{equation}

where $w_{r}$ and $w_{z}$ = $S\times w_{r}$ are the radial and axial waists of the confocal PSF, assumed to be a 3D Gaussian function. The diffusion time, $\tau_{D}$, is related to the radial waist, $w_{r}$, and to the diffusion constant, $D$, by:

\begin{equation}
\tau_{D} = \dfrac{w_{r}^{2}}{4D}
\label{Tau-D}
\end{equation}

Let us recall that, for a given temperature, the diffusion constant is inversely proportional to the viscosity (Stoke-Einstein relation) and that the diffusion time is therefore proportional to the viscosity.

In addition to diffusion that causes fluctuations at relatively long times, intramolecular dynamics between the singlet and triplet states modifies the short time behavior of the ACF. This short time behavior is characterized by the triplet lifetime, $\tau_{T}$ and the triplet fraction, $f_{T}$, that is the average fraction of molecules in the triplet state \cite{Triplet}. Altogether, this leads to the following model to fit the ACF:

\begin{equation}
G(\tau) = 1 + \dfrac{1}{N}\left(1 + \dfrac{f_T}{1 - f_T}e^{-\tau/ \tau_T}\right)\left(1 + \dfrac{\tau}{\tau_D}\right)^{-1}\left(1 + \dfrac{\tau}{S^2 \tau_D}\right)^{-1/2}
\label{GdeTau}
\end{equation}

At this point it is interesting to note that, among the FCS parameters, the number of molecules, $N$, is the less model dependent one. The reason is that, for time lag tending to zero, the amplitude of the ACF is proportional to $1/N$, the proportionality factor depending only on the triplet fraction, $f_{T}$, which is, for a fixed laser power, constant.

In this article we neglect photophysical effects, such as optical saturation or coupling between excitation and emission dipole orientations \textit{etc.}, so that the Molecular Detection Efficiency function, used in FCS \cite{hess02}, can be equated to the confocal PSF. Let us recall that the later is, by definition, the 3D image that would be reconstructed after a 3D scanning in the vicinity of a fluorescent nanosphere. It will be approximated as the product of the illumination and detection PSFs, because we suppose an infinitesimally small detector. In addition, by analogy with the Molecular Detection Efficiency function, the intensity of the confocal PSF is normalized to 1 at the origin. According to the theory of FCS, the general relation between the confocal PSF, $PSF_{con}$, and the FCS observation volume, $V_{fcs}$, reads \cite{TheoryFCS}:

\begin{equation}
V_{fcs} = \dfrac{[\int PSF_{con}d\vec{r}]^{2}}{\int PSF_{con}^{2}d\vec{r}}
\end{equation}

The confocal PSF having a 3D Gaussian profile, the FCS observation volume satisfies $V_{fcs} = \pi^{3/2}{w_{r}^{2}}{w_{z}}$, which is consistent with Eq. \ref{NCWrWz}. In a confocal configuration, with an homogeneous solution taken as sample (or with a very weakly contrasted object), the total Count Rate, $CR$, is proportional to the integral of the product of the illumination intensity profile times the collection efficiency profile. In presence of aberrations, both illumination and collection profiles are expected to show a lower peak value and enlarged width. The Strehl ratio accounts, by definition, for the attenuation of the peak intensity. Introducing $S_{tr}$ as the single-pass Strehl ratio, which is assumed to have the same value for the illumination and collection paths \cite{Strehl}, we can write the count rate as:

\begin{equation}
CR = \eta\times C\times S_{tr}^{2}\int PSF_{con}d\vec{r}
\end{equation}

 $\eta$ takes into account the laser power in the sample, the absorption cross section, the fluorescence quantum yield and the overall detection efficiency (including lenses, fluorescence filters and efficiency of the photon detector). Note that, still under the assumption of a Gaussian profile, the integral of the confocal PSF reads $(\pi /2)^{3/2}\times{w_{r}^{2}}{w_{z}}$. Therefore, the molecular brightness, defined as the Count Rate per Molecule, that is $CRM = CR / (C\times V_{fcs})$, reads:

\begin{equation}
CRM = \dfrac{\eta}{2^{3/2}}S_{tr}^{2}
\label{crm}
\end{equation}

A strong consequence of the assumed Gaussian profile for the PSF is thus that the molecular brightness depends only upon the Strehl ratio. In addition, we numerically checked for the aberrations considered in this paper that the Strehl ratio can be very well approximated by:

\begin{equation}
S_{tr}\simeq \exp ^{ -\left(\dfrac{2 \pi \times \sigma_{wf}}{\lambda}\right)^{2}}
\label{strehl}
\end{equation}

$\sigma_{wf}$ is the Root Mean Square (RMS) amplitude of the aberrations (of both the illumination and detection beams) and $\lambda$ is the mean value of the excitation and fluorescence wavelengths. In practice, the amplitudes of Zernike aberrations introduced by the sample can be estimated, by assuming that the aberrations are perfectly corrected by the DM. For each Zernike mode $i$, we note $a_{i}$ the amplitude of the aberration mode $i$. Piston, tilts and defocus terms are ignored in our work, because, to first order, they have no impact on the shape of the confocal PSF. Thanks to the orthonormal property of Zernike aberrations, the overall RMS amplitude of all the modes reads:

\begin{equation}
 \sigma_{wf}=\sqrt{\sum_{i=4}^{i=10} a_{i}^2}
\label{rms}
\end{equation}

Aberration corrections were performed for seven Zernike modes: astigmatisms ($a_{4}$, $a_{5}$), comas ($a_{6}$, $a_{7}$), trefoils ($a_{8}$, $a_{9}$), and primary spherical aberration ($a_{10}$). Our analysis of the impact of aberrations on the estimated FCS parameters ignored other Zernike modes, but nevertheless describes our experimental results accurately.

\section{Experimental setup and materials}
\subsection{Adaptive Optics}

\begin{figure}[htbp]
\centering\includegraphics[width=12cm]{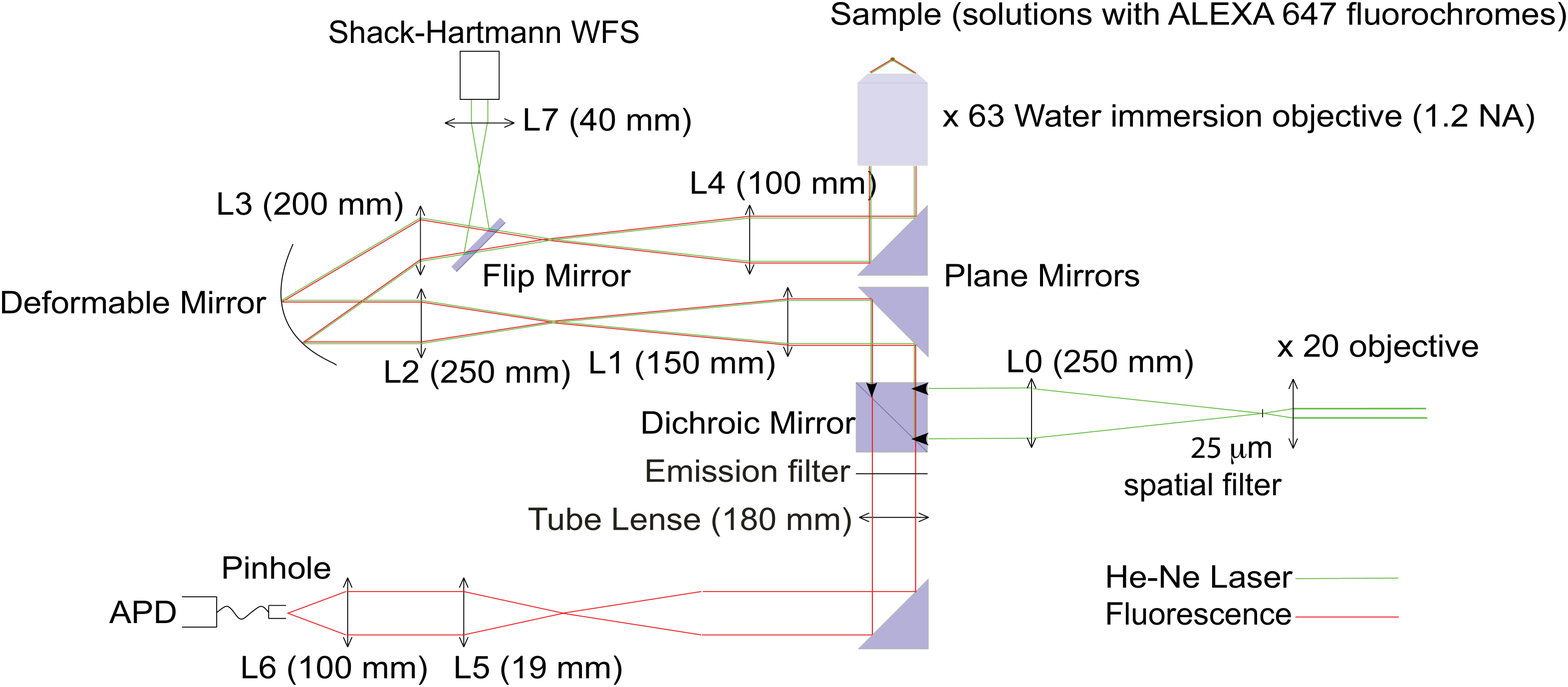}
\caption{Optical layout of the experiment; the flip mirror is removed while performing FCS measurements.}
\label{System}
\end{figure}

The optical system of our experiment was built on an Olympus X71 platform using a \mbox{ C-Apochromat $63 \times /1.2$ WKorr} water objective manufactured
by Zeiss and a fluorescence cube with a dichroic mirror and an emission filter (z633rdc and HQ700/75m from Chroma). We show its schematic in Fig. \ref{System}. A non-scanning confocal detection arm was built using the left port of the platform and consists of a point
detector (multimode fiber with a 50 $\mu$m core radius coupled to an Avalanche Photodiode (APD) from PerkinElmer) and a $\times 5.3$ imager (lenses $L5$ and $L6$). The
overall magnification of this confocal microscope  being 276, the detector radius corresponds to 2.15 optical units. The laser beam at 633 nm (HeNe, from Thorlabs) has a uniform intensity profile in the pupil plane of the objective, which is required to have a uniform signal to noise ratio on the Shack-Hartmann Wavefront Sensor (SHWFS) when performing the calibration of the DM. 

\begin{figure}[htbp]
\centering\includegraphics[scale=0.4]{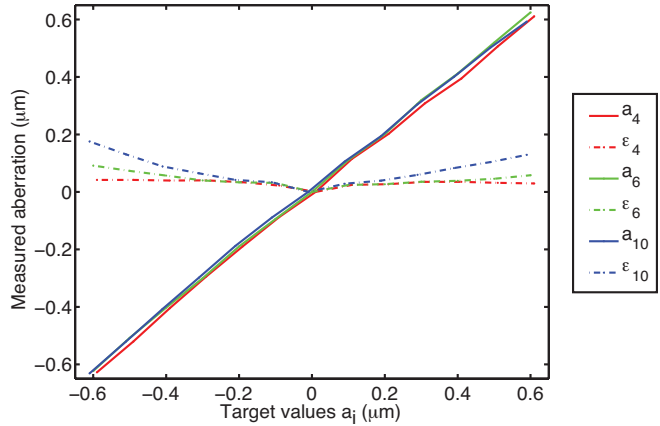}
\caption{Calibration of the DM with a SHWFS. The amplitudes of the measured Zernike modes (solid lines), astigmatism ($a_{4}$), coma ($a_{6}$) and spherical aberration ($a_{10}$) are plotted versus the targeted values (x-axis). The corresponding RMS residual errors $\epsilon_{4}$, $\epsilon_{6}$, $\epsilon_{10}$. are alos shown (dash dot lines)}
\label{SHWFS}
\end{figure}

The Hi-speed 97 DM (manufactured by ALPAO) has the ability to generate  with good precision Zernike aberrations of large amplitudes. The SHWFS (manufactured by ALPAO) analyses the laser beam thanks to a flip mirror located between the lenses L3 and L4. We recall that the use of a SHWFS for
aberration measurements in a microscope is usually prevented by its lack of optical sectioning capability \cite{booth09}, unless the object is labelled
with fluorescent microspheres \cite{azucena1,azucena2,tao1,tao2}. Henceforth, direct control of the DM with the SHWFS is not possible during a FCS experiment. The 13.75 mm pupil of the DM is magnified by a factor 0.25 on the SHWFS using a
telescope (L3, L7) so that the measured wavefront is sampled by 32 microlenses across the pupil diameter. This calibration, performed using MATLAB
built-in functions provided by ALPAO, makes it possible to generate Zernike aberrations in open-loop. We tested the accuracy of this open-loop
control using the SHWFS, and we found that the measured  aberrations were very close to the target aberrations. For each Zernike mode~$i$ of amplitude $a_i$, one can quantify the
departure from the target aberration using the RMS error  of the open-loop, $\epsilon_{i}$, which is measured by the SHWFS, using 75 Zernike modes. Results are shown in Fig. \ref{SHWFS} for astigmatism $a_{4}$, coma $a_{6}$, and spherical aberration $a_{10}$. The error is smaller than 25 nm in the range of aberrations that we corrected in the present study ($|a_{10}|<0.1~\mu$m). It is worthwhile noting that the error is larger for spherical aberration, because the linear model of the open-loop DM control
is less accurate for higher order aberrations.

Prior to each FCS experiment, we performed aberration corrections with the so called sensorless AO system \cite{neil00,booth02}: we optimized each Zernike mode sequentially using the mean count rate of photons detected by the APD during 1 second. We took three measurements per mode $i$, with the DM biased at $a_{i}=-0.1, 0, 0.1~\mu$m. A parabolic interpolation of the corresponding measurements of the count rate yielded the optimal amplitude $\widehat{a_{i}}$ that will be retained for each mode $i$. Final correction was obtained after cycling twice through all the 7 Zernike modes and thus took 42 s.

\subsection{Fluorescence excitation, data acquisition and treatment}
The laser power within the sample was set to 30 $\mu W$, in order to avoid any saturation effect that would affect the shape of the PSF. In case of no aberration, this leads to a typical count rate per molecule of 7 kHz. The digital signal of the APD was sent to an homemade data acquisition system based on a PCI 6602 card from National Instrument, which provides real time evaluation of the count rate and raw data saving. Each ACF curve, with its mean value and standard error of the mean, was obtained by performing 10 acquisitions of 10 seconds. Raw data processing (ACF calculations and fits) were performed using MatLab (Mathworks) and Origin (OriginLab Corp.). Since, in case of aberrations (\textit{i.e.} for glycerol solutions), the value of the structure parameter, $S$, is ill defined, the corresponding fits were performed by setting $S$ to its estimated value after aberration corrections (typically found between 5 and 7).

\subsection{Fluorophores}
All our experiments have been performed at $22^{\circ}C$, with Alexa Fluor 647 (A647), purchased from Invitrogen Molecular Probe. Stock solutions were prepared without further purification. The molecular concentration of the A647 solutions was 80 nM, unless specified.
We used A647 in pure water and in two aqueous solutions of glycerol: 50$\%$ (v/v) and 70.4$\%$ (v/v). Using a rheometer (Anton Paar, model MCR 301), we controlled the viscosities of the glycerol solutions and found 7.86 mPa.s for the 50$\%$ glycerol solution and 30.7 mPa.s for the 70.4$\%$ one. Since these values are in good agreement with the tabulated values of water-glycerol mixtures \cite{eta-Gly}, we can confidently estimate the refraction indices to their tabulated values: 1.407 for the 50$\%$ glycerol solution and 1.435 for the 70.4$\%$ one \cite{index-Gly-eta-water}.

\section{Experimental results and discussion}

\subsection{Calibration at the cover slide - sample interface}

Prior to the experiments with the glycerol solutions, we have corrected the aberrations of our optical system with an A647 80 nM pure water solution. The main aberration was astigmatism ($a_{5}\simeq0.2~\mu$m), which is probably introduced by the dichroic mirror and the surface of the DM that is non-flat when no commands are applied. The correction collar of the objective was also manually optimized before starting aberration corrections and then set to this adjustment for the entire experiment. These optimized set of Zernike  aberrations was then systematically applied to the DM before any FCS measurement. Henceforth, in the case of glycerol solutions, the outcome of our experiments only depends on whether we corrected the remaining aberrations introduced by the sample or not. The reference depth was taken at $z = 10 \mu m$, to avoid any artifact due to the cover slide - solution interface.

 The diffusion constant of A647 in pure water has been precisely measured elsewhere at $25^{\circ}$C by dual focus FCS \cite{picoquant}. Making a slight correction to account for our experimental temperature ($22^{\circ}C$) \cite{index-Gly-eta-water}, we derive a reference value $D_{water} = 304.5 \mu m^2/s$. From the measured diffusion time of A647 in pure water, $\tau_D = 48 \mu s$ (data not shown), we use Eq. \ref{Tau-D} to deduce the radial width of the confocal PSF and found $w_{r} = 0.242 \mu m$. In addition, using the measured structure parameter and number of molecules, $S = 10$ and $N = 34.6$, we obtain, using Eq. \ref{NCWrWz}, a concentration of 73 nM, in good agreement with the one we intended to prepare (80 nM). Our measured diffusion time of A647 in pure water can now be compared with the corresponding values in aqueous solutions of glycerol. Using the viscosities of water at $22^{\circ}C$ (0.955 mPa.s \cite{index-Gly-eta-water}) and of our glycerol solutions (7.86 and 30.7 mPa.s, see above), we derive a diffusion time $\tau_D = 395 \mu s$ in the 50$\%$ glycerol solution and $\tau_D = 1545 \mu s$ in the 70.4$\%$ one. Of course, these derivations assume that the confocal radial waist, $w_{r}$, is the same in water and in the glycerol solutions, so that the diffusion time is proportional to the viscosity. In other words, the calculated diffusion times in the aqueous solutions of glycerol correspond to perfectly corrected optical aberrations. Experimentally, we measured, at the reference depth of $z = 10 \mu m$,  $\tau_D = 410 \mu s$ for the 50$\%$ glycerol solution (data not shown) and $\tau_D = 1391 \mu s$ for the 70.4$\%$ one (see the corresponding ACF in Fig. \ref{ACFs}, red curve in the left graph). The 10$\%$ discrepancy of $\tau_D$ from its expected value, in the case of the 70.4$\%$ glycerol solution, can be attributed to the non perfect aberration corrections at the interface.

\subsection{Measurements as a function of the focussing depth}

\begin{figure}[h!]
\centering\includegraphics[width=10cm]{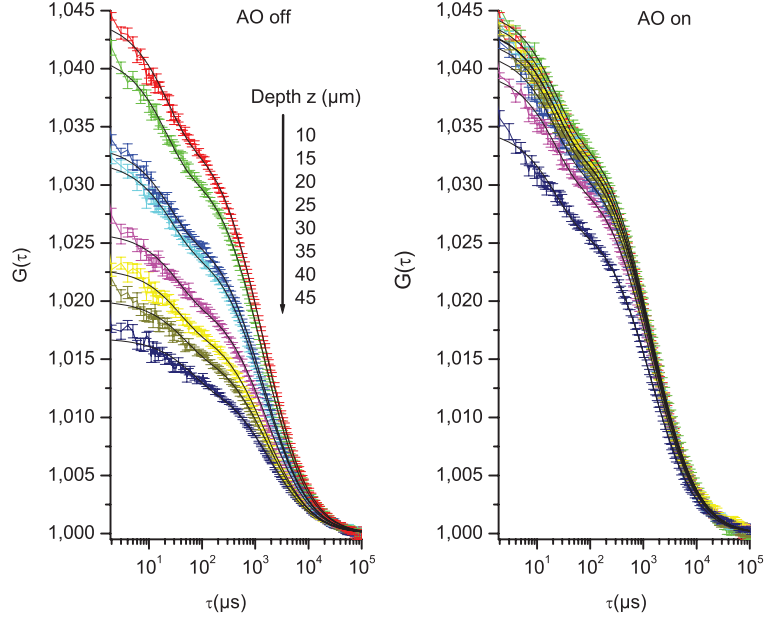}
\caption{ACF recorded in the 70.4 $\%$ glycerol solutions, without (left) and with (right) AO. The amplitude of the ACF decreases dramatically with increasing observation depth (from 10 to 45 $\mu m$) when the AO is not switched on. The superimposed dark solid lines are the fits performed with Eq. \ref{GdeTau}}
\label{ACFs}
\end{figure}

As anticipated, optical aberrations have a drastic impact on the ACF data. Fig. \ref{ACFs} shows, at first glance, the systematic decrease in the amplitude of the ACF when we focus in the $70.4\%$ glycerol solution (left graph). This corresponds to an increase in the number of molecules, $N$ (Eq. \ref{GdeTau}) due to the increase in the FCS observation volume (Eq. \ref{NCWrWz}). Our AO system provides an efficient aberration correction, and the differences between the ACF curves are greatly reduced (right graph). The AO correction was clearly less efficient at $z = 45~\mu m $ focussing depth and we therefore did not acquire data deeper into the $70.4\%$ glycerol solution. In contrast, with the $50\%$ solution, we obtained an efficient aberration correction down to $z = 80 \mu$m (data not shown). This is because the refraction index mismatch is lower with this solution.

\begin{figure}[h!]
\centering\includegraphics[width=12cm]{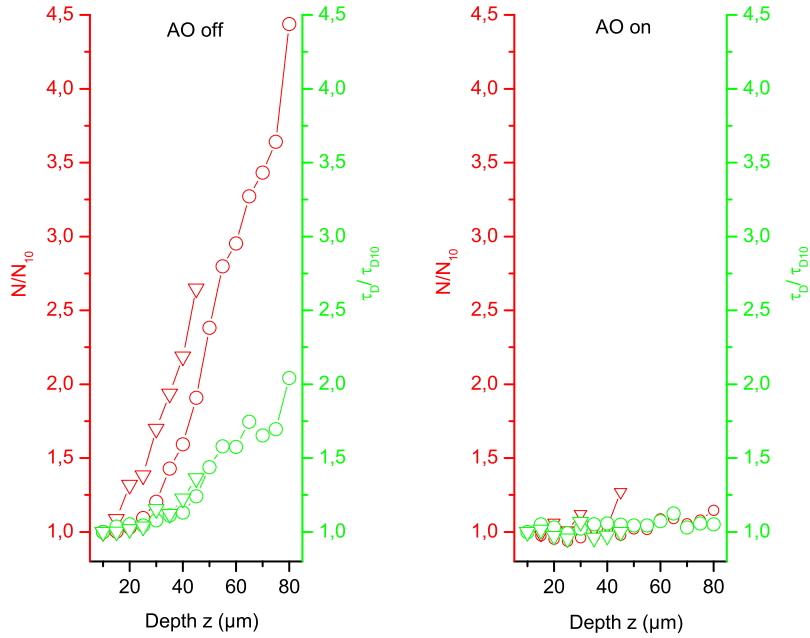}
\caption{Variations of the estimated FCS parameters normalized to their values at the reference depth $z = 10 \mu m$, $N/N_{10}$ (red) and $\tau_D/ \tau_{D10}$ (green), in the 70.4 $\%$ (open triangles) and 50 $\%$ (open circles) glycerol solutions, without (left) and with (right) AO.}
\label{fcsfit}
\end{figure}

Using the analytical expression for the ACF (Eq. \ref{GdeTau}), we fit the ACF data to estimate $N$ and $\tau_{D}$ at each focussing depth, for the two glycerol solutions. Fig. \ref{fcsfit} show results that are normalized to the values obtained at the reference depth of $z = 10 \mu m$, without and with AO switched on (left and right graph respectively). Without AO, $N$ and $\tau_{D}$ are more sensitive to the focussing depth with  the $70.4\%$ glycerol solution, because the larger refraction index mismatch the larger the aberrations. Note that $\tau_{D}$ depends only upon the lateral size of the confocal PSF (Eq. \ref{Tau-D}), while $N$ depends upon the volume (Eq. \ref{NCWrWz}), which explains why the relative increase of $N$ is more pronounced than that of $\tau_{D}$. The right panel of Fig. \ref{fcsfit} exemplifies the very efficient aberration correction.

\clearpage

\begin{figure}[h!]
\centering\includegraphics[scale=0.4]{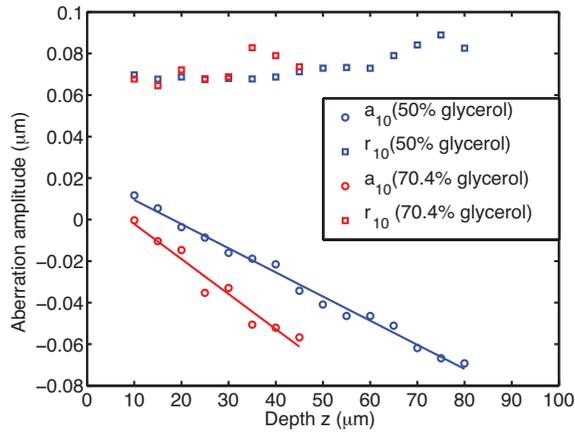}
\caption{Aberrations corrected by the DM in the glycerol solutions: spherical aberration amplitudes ($a_{10}$, open circles) and the residual aberrations ($r_{10}$, open squares) for the 50$\%$ glycerol solution (blue) and for the 70.4$\%$ one (red); solid lines are linear fits of the spherical aberration amplitudes.}
\label{modescorriges}
\end{figure}

Assuming that the AO system performs a perfect aberration correction, we can estimate the aberrations introduced by the sample looking at the value of the optimized Zernike modes. We show in Fig. \ref{modescorriges} the amplitudes of spherical aberrations ($a_{10}$, open circles) and residual RMS of all the other modes ($r_{10}$, open squares) generated by the DM for the two glycerol solutions. As expected, the refractive index mismatch introduces a spherical aberration, the amplitude of which is proportional to the focussing depth \cite{booth98}. In addition, the slope of the corresponding linear fit is 1.5 times larger in the 70.4$\%$ glycerol solution that has a higher refractive index mismatch.

Using the measured molecular brightness, $CRM$ and the amplitude of the Zernike aberrations, we can compare the experimental data with the modeling presented in Section 2, which shows that the molecular brightness scales as the Strehl ratio squared (see Eq 7).

With the glycerol solutions, we compute the overall RMS aberrations introduced by the sample as the quadratic sum of the optimized Zernike modes, with the $z=10\mu$m depth taken as reference: $\sigma_{wf}(z)=\sqrt{\sum_{j=4}^{j=10}[a_{j}(z)-a_{j}(10)]^{2}}$. We plot in Fig. 6 (left panel) the Strehl ratio squared, $S_{tr}^{2}$, function of $\sigma_{wf}$ (black solid line), using Eq. 8 and the brightness obtained in the two glycerol solutions. Brightness values are computed using the fit values of $N$ (Fig.4, left panel) and the corresponding count rate, $CR$ (data not shown) without AO correction. They are normalized with the value measured at $z=10\mu$m. The agreement between experimental data and the predictions of Eq. 7 is good, even though the Gaussian approximation of an aberrated confocal PSF is probably inaccurate.

\begin{figure}[htbp]
\centering\includegraphics[scale=0.35]{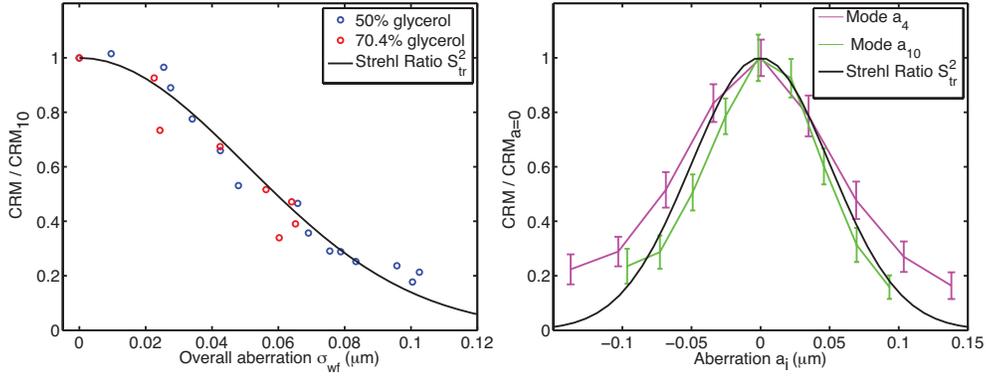}
\caption{Comparison of the molecular brightness with the Strehl ratio squared, computed using Eq. \ref{strehl} (solid lines). Left graph: the ordinate on the vertical axis, $CRM/CRM_{10}$, is normalized to its value at the reference depth $z = 10 \mu m$, the horizontal axis is the overall RMS amplitude of the corrected aberrations, $\sigma_{wf}$, in the glycerol solutions (50 $\%$ solution in blue and 70.4 $\%$ one in red). Right graph: the ordinate on the vertical axis, $CRM/CRM_{a=0}$ is normalized to its value when no single mode aberration is applied, the horizontal axis is the amplitude of a single Zernike mode ($a_{4}$: astigmatism in magenta, $a_{10}$: spherical aberration in green), generated by the DM in an A647 80 nM pure water solution, while other aberrations are corrected.}
\label{CRM}
\end{figure}

Similarly, we confront our modeling to data obtained in the A647 80 nM pure water solution. For this experiment, a single Zernike mode was intentionally introduced by the DM, after the AO system had corrected the system aberrations. In this case, the overall RMS aberrations $\sigma_{wf}$, used to compute the Strehl ratio squared (with Eq. 8), equates to the amplitude of the single Zernike mode, $a_{i}$. We performed the experiments for Zernike astigmatism $a_{4}$ and spherical aberration $a_{10}$. The molecular brightness, normalized with the value measured for $a_{i}=0$, is shown in Fig. 6 (right panel). We observe a good agreement with the model, although there is a slight
discrepancy for astigmatism. Two reasons could explain that the modeling is less accurate with the astigmatism mode: i) the image of the microscope objective exit pupil in the plane of the DM is elongated, because of a $15^{\circ}$ incidence angle on the DM; ii) the lack of radial symmetry of the confocal PSF when astigmatism is introduced, which cannot be taken into account with the standard ACF modeling of Eq. 4. Curves very similar to the ones of the right panel of Fig.6 have been obtained with a A647 8 nM pure water solution (data not shown).

We have shown in various experimental conditions (glycerol proportions, types of Zernike aberrations, fluorophore concentrations) that the model of Eq.7 can be used as a simple rule of thumb for the design of an AO system for FCS, or to anticipate the impact of aberrations in FCS measurements.

\section{Conclusion}
We have shown that a moderate refraction index mismatch ($\Delta n$ up to 0.1) can have a dramatic impact on FCS data, when focusing at a few tens of $\mu m$ above the cover slide - solution interface. Although the wavefront is weakly distorted (aberration amplitude has a RMS smaller than 0.1 $\mu m$), the FCS parameters are strongly biased (the diffusion time is multiplied by up to a factor 2 and the number of molecules by more than a factor 4). Such effects constitute a technical bottleneck, since an ideal FCS experiment requires a perfectly controlled observation volume, in order to compare FCS parameters obtained with different samples of interest. We showed that Adaptive Optics makes it possible to stabilize the observation volume in solutions of fluorescent molecules of nM concentrations. Interestingly , the count rate per molecule (or molecular brightness), as provided by FCS, scales as the square of the Strehl ratio. It is remarkable that, thanks to FCS, this key quantity can be obtained without acquiring an image. We demonstrated this idea in homogeneous media, but it could be extended to weakly contrasted samples. Thus, we suggest that the count rate per molecule could be used as a optimization metric when applying Adaptive Optics to biological media. More importantly for biological applications of FCS, after aberration corrections, the count rate per molecule would more confidently reflects environmental conditions, such as fluorescent probe aggregation, pH variations, \textit{etc}. In the near future, our AO system should improve significantly the robustness of FCS measurements in environments of various optical properties (crowded solutions, cellular media, tissues, etc.).

\section{Acknowledgements}
The authors thanks F. Rooms and S. C\`{e}tre from ALPAO for valuable technical assistance with the Deformable Mirror, C. Schneider for data manipulation and G. Coupier for the viscosity measurements. C.-E. L. thanks the CNRS for the post-doctoral position.

\end{document}